\documentclass[journal,10pt,twocolumn]{IEEEtran}
\usepackage{epsfig,fullpage,amsmath,cite}
\usepackage{amssymb,multirow,delarray}
\usepackage{epsfig}
\usepackage{graphicx}
\usepackage{subfigure}

\newtheorem{proposition}{Proposition}

\newtheorem{definition}{Definition}

\begin{document}

\title{\huge{High Performance Cooperative Transmission Protocols Based on Multiuser Detection and Network Coding}}
\author{Zhu Han$^1$, Xin Zhang$^2$, and H. Vincent Poor$^3$\\
$^1$Department of Electrical and Computer Engineering,
 Boise State
University, Idaho, USA\\
$^2$The United Technologies Research Center,  411 Silver Lane,
East
Hartford, Connecticut, USA\\
$^3$School of Engineering and Applied Science,  Princeton
University, New Jersey, USA. \vspace{-7mm} \thanks{This research
was supported by the National Science Foundation under Grants
ANI-03-38807 and CNS-06-25637.}}

\maketitle
\begin{abstract}
Cooperative transmission is an emerging communication technique
that takes advantage of the broadcast nature of wireless channels.
However, due to low spectral efficiency and the requirement of
orthogonal channels, its potential for use in future wireless
networks is limited. In this paper, by making use of multiuser
detection (MUD) and network coding, cooperative transmission
protocols with high spectral efficiency, diversity order, and
coding gain are developed. Compared with the traditional
cooperative transmission protocols with single-user detection, in
which the diversity gain is only for one source user, the proposed
MUD cooperative transmission protocols have the merit that the
improvement of one user's link can also benefit the other users.
In addition, using MUD at the relay provides an environment in
which network coding can be employed. The coding gain and high
diversity order can be obtained by fully utilizing the link
between the relay and the destination. From the analysis and
simulation results, it is seen that the proposed protocols achieve
higher diversity gain, better asymptotic efficiency, and lower bit
error rate, compared to traditional MUD schemes and to existing
cooperative transmission protocols. From the simulation results,
the performance of the proposed scheme is near optimal as the
performance gap is $0.12$dB for average bit error rate (BER)
$10^{-6}$ and $1.04$dB for average BER $10^{-3}$, compared to two
performance upper bounds.

\end{abstract}


\section{Introduction}\label{sec:intro}

Cooperative transmission \cite{bib:Aazhang1, bib:Laneman2} takes
advantage of the broadcast nature of wireless channels to improve
data transmission through cooperation among network nodes.
Notably, relay nodes can be employed as  virtual antennas for a
source node, so that the multiple input multiple output (MIMO)
technology can be exploited even with single-antenna terminals.
Recent work has explored cooperative transmission in a variety of
scenarios, including cellular networks \cite{bib:ICC06}, ad
hoc/sensor networks \cite{Luo,bib:WCNC06_sensor,bib:Madsen},
WiFi/WiMax \cite{bib:WCNC06_WIFI} and ultra-wideband
\cite{bib:WCNC06_UWB}. One drawback of existing cooperative
transmission schemes is a consequent reduction of spectral
efficiency due largely to the fact that most such techniques
require orthogonal channels for the transmissions of cooperating
nodes\footnote{It is worth mentioning that the spectral efficiency
in the cooperative transmission literature is defined as the
number of orthogonal channels required for direct transmission
divided by the overall number of channels for both direct
transmission and relaying. This definition is different from that
of spectral efficiency typically used in the adaptive modulation
literature \cite{Goldsmith}.}. This requirement is limiting, as
many wireless networks, such as 3G cellular networks, cannot
provide orthogonal channels.

In this paper, we consider cooperative transmission protocols for
networks that do not require orthogonality among the signaling
channels of the nodes in the network.  Such a scenario naturally
motivates the use of multiuser detection (MUD) \cite{verdu} to
mitigate the interference caused by non-orthogonal signaling. The
performance of MUD is generally good when interfering users have
significantly different link conditions from one another. In
traditional MUD, the link conditions are determined by users'
locations and channel gains, which are not controllable by the
designer. However, with cooperative transmission, we have the
opportunity to  optimize such conditions by deciding which relay
will retransmit which user's information so that the  selected
users' link conditions can be optimized for overall system
performance. A link level analysis for MUD over cooperative
transmission can be found in \cite{Venturino}.

Recently \cite{HuaiyuDai} has considered the joint optimization of
MIMO systems with MUD. However, unlike MIMO MUD in which all
information from different antennas can be obtained without
limitation, in cooperative communications the information
transmission between the relay (i.e., the virtual antenna) and the
destination is restrained by a lossy relay-destination wireless
link. To overcome this limitation, network coding
\cite{network_coding, network_coding_coop} provides a potential
solution. The core notion of network coding is to allow mixing of
data at intermediate network nodes to improve the overall
reliability of transmission across the network. A destination
receives these coded data packets from various nodes and deduces
from them the messages that were originally intended for that
destination. In \cite{CISS}, it is seen that information exchange
can be efficiently performed by exploiting network coding and the
broadcast nature of the wireless medium. In cooperative
transmission, the relay can be viewed as an intermediate network
node. In \cite{Laneman}, the network coding gains of various
cooperative diversity protocols are examined in detail. In this
paper, we consider the situation in which  MUD is employed at the
relays, so that a  relay can obtain information from various users
and then use network coding by mixing multiple users' data and
transmitting coded information through the limited
relay-destination link. In other words, MUD provides an
environment for deploying network coding, and network coding can
achieve substantial coding gain and high diversity to overcome the
limitations of the relay-destination link.

In particular, we propose two cooperative transmission protocols
that utilize MUD and network coding. In the first protocol,
realizing that improvement in one user's detection can help the
detection of the other users in certain types of multiuser
detectors (e.g., interference cancelers), we decide which relays
to use and whose information the selected relays will retransmit
such that the overall system performance can be optimized at the
sink node. In the second protocol, we assume the relays are
equipped with MUD. Then the selected users' information is coded
by network coding and is relayed to the base station. At the base
station, the coding gain is not only realized for the selected
users but also for the other users because of MUD. Moreover, we
develop two performance upper bounds to evaluate the proposed
schemes. Practical implementation issues are also discussed. From
both analytical and simulation results, it is seen that the
proposed protocols achieve higher diversity and coding gains,
better asymptotic efficiency, and lower bit error rate (BER) than
existing schemes without sacrificing spectral efficiency. The
proposed scheme achieves performance less than 0.12dB away from
the performance upper bounds when the average BER equals
$10^{-6}$, and 1.04dB when BER equals $10^{-3}$.

This paper is organized as follows: In Section \ref{sec:model},
system models are given for cooperative transmission and MUD in a
network consisting of a source node (e.g., a mobile terminal), a
sink node (e.g., a base station or access point)
 and a set of relays. In Section \ref{sec:Protocol},
the two above-mentioned protocols are constructed. In Section
\ref{sec:analysis}, the properties of the proposed protocols are
analyzed. Simulation results are shown in Section
\ref{sec:simulation}, and conclusions are drawn in Section
\ref{sec:conclusion}.

\begin{figure}[htbp]
\begin{center}
    \includegraphics[width=80truemm]{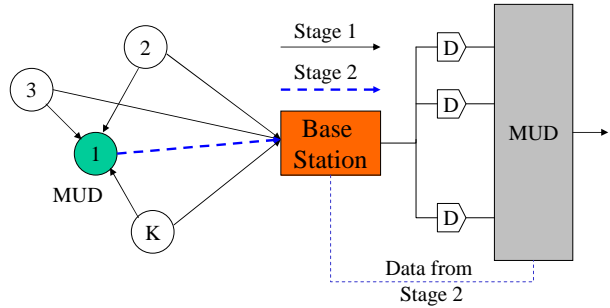}
\end{center}
\caption{Proposed Cooperative System model}\label{system_model}
\end{figure}

\section{System Model}\label{sec:model}

We consider an uplink synchronous code-division multiple-access
(CDMA) system with Gaussian ambient noise\footnote{Note that
asynchronous CDMA can be treated similarly.}. There are  $K$
synchronous uplink users (i.e., terminals) each with a single
antenna. Here the number of users is no more than the number of
available CDMA codes. Among these terminals, $N$ can serve as
relays. This system model is illustrated with $N=1$ in Figure
\ref{system_model}. At the first transmission stage, all users
except the relays send information, and the relays listen (and
perform MUD if they have the ability). At the second stage, the
other users send their next information signals, while the relays
send a certain user's information or the networking-coded
information from the results of MUD applied at the first stage. In
the sink node, which for convenience we will refer to as the  a
base station, all of the other users' information from the first
stage is delayed by one time slot and jointly detected with the
information sent by the relays at the second stage. Since the
users cannot transmit and receive at the same time or on the same
frequency, to relay once costs at least two time slots for
listening and relaying. So the spectral efficiency is $\frac {K-N}
K$, and thus when the number of users is much larger than the
number of relays the spectral efficiency approaches one. On the
other hand, in the traditional cooperative transmission with
one-relay and one-source pair, $N=\frac K 2 $. In this case, the
spectral efficiency is $\frac 1 2$.

We denote by $\mathfrak{R}$ the group of relay terminals, and by
$\mathfrak{L}$  the group of terminals that are listening and will
serve as relays in the next time slot\footnote{This is because of
half duplex.}. Define the set $\mathfrak{K}$ for all $K$ users. In
the first stage, the received signal at the base station can be
expressed as
\begin{equation}
y(t)=\sum_{k\in \mathfrak{K}\backslash \mathfrak{R}  \backslash
\mathfrak{L} } A_k b_k s_k(t)+ \sum_{k\in  \mathfrak{R}  }
A_kz_ks_k(t)+ \sigma n(t),
\end{equation}
and at user $i\in\mathfrak{L}$, who is listening and preparing for a
relay in the next time slot, as
\begin{equation}
y^i(t)=\sum_{k\in \mathfrak{K}\backslash \mathfrak{R}  \backslash
\mathfrak{L} } A_k^i b_k s_k(t)+ \sum_{k\in  \mathfrak{R}  } A_k^i
z_k s_k(t)+\sigma ^i n^i(t),
\end{equation}
where $A_k$ is the received amplitude of the $k^{th}$ user's
signal at the base station, $A_k^i$ is the received amplitude of
the $k^{th}$ user's signal at relay $i$, $b_k\in \{ -1,+1\}$ is
the data symbol transmitted by the $k^{th}$ user, $z_k$ is the relayed
bit, $s_k$ is the unit-energy signature waveform ( i.e.
Pseudo-random code) of the $k^{th}$ user, $n(t)$ and $n^i(t)$ are
the normalized white Gaussian noise, and $\sigma^2$ and $(\sigma
^i)^2$ are the background noise power densities. For simplicity,
we assume $\sigma=\sigma^i$, although the more general case is
straightforward.

The received signal vectors at the base station and at the relay
after  processing by a  matched filter bank can be written as
\begin{equation}\label{A}
\textbf y=\textbf{R}\textbf{A}\textbf{b}+\textbf n,
\end{equation}
and
\begin{equation}
\textbf y ^i=\textbf{R}\textbf{A}^i\textbf{b}+\textbf n ^i,
\end{equation}
where $\textbf R$
is the signal cross-correlation matrix, whose elements can be written
as
\begin{equation}
    [\textbf R]_{ij}=\int _{0}^T s_i(t)s_j(t)dt,
\end{equation}
with $T$  the inverse of the data rate, $\textbf A=\mbox{diag}\{A_1,\dots , A_{K}\}$, $\textbf
A^i=\mbox{diag}\{A^i_1,\dots , A^i_{K}\}$, $E[\textbf n \textbf
n^T]=E[\textbf n^i \textbf n^{iT}]=\sigma^2\textbf R$, and $\textbf b=[b_1,
\dots, z_i, \dots, 0, b_{K}]^T$  consists of symbols of
direct-transmission, relay, and listening users. In particular,  $b_i$ is the
direct-transmission symbol, $z_i$ is the relay symbol, and the
listening relay has zero to transmit due to the half duplex
assumption.

From the cooperative transmission perspective, in the first stage
user $k\in \mathfrak{K}\backslash \mathfrak{R} \backslash
\mathfrak{L}$ transmits its signal directly to the base station,
and user $i \in \mathfrak{L}$ listens. In the second stage, the
users listening in the first stage become relays (set
$\mathfrak{R}$) and relay the information to the base station. At
the base station, the information at the first stage is delayed by
one time slot and then is combined with the information at the
second stage.

In this paper, we will investigate the BER performance of MUD
under cooperative transmission. Specifically, we will consider
optimal MUD and the successive cancellation detector, which
is one type of decision-driven MUD.

As pointed out in \cite{verdu}, there is no explicit expression
for the error probability of the optimal multiuser detector, and
bounds must be used. A tight upper bound is provided by the
following proposition from \cite{verdu}.
\begin{proposition}
The BER of the $i^{th}$ user for optimal MUD is given
by
\begin{equation}\label{BER_Opt}
P_r^{i,opt}\le \sum_{\mathbf{\epsilon} \in F_i}
2^{-\omega(\mathbf{\epsilon})} Q\left(
\frac{\|S(\mathbf{\epsilon})\|}{\sigma} \right)
\end{equation}
where $\mathbf{\epsilon}$ is a possible error vector for user $k$,
and $\|S(\mathbf{\epsilon})\|^2
=\mathbf{\epsilon}^T\mathbf{H}\mathbf{\epsilon}=
\mathbf{\epsilon}^T\mathbf{ARA}\mathbf{\epsilon}$.
$\omega(\mathbf{\epsilon})$ is the number of nonzero elements in
$\mathbf{\epsilon}$, and $F_i$ is the subset of indecomposable
vectors. (See \cite{verdu} for details.)
\end{proposition}

For the successive cancellation detector, a recursive
approximation for the error probability
 is given by the following
proposition\cite{verdu}.
\begin{proposition}
The BER of the $i^{th}$ user for successive cancellation is given approximately by
\begin{equation}\label{BER_SC}
P_r^{i,sc}\approx Q \left( \frac {A_i}{\sqrt{\sigma^2+\frac 1 M
\sum_{j=1}^{i-1}A_j^2+\frac 4 M \sum_{j=i+1}^K
A_j^2P_r^{j,sc}}}\right) ,
\end{equation}
where $M$ is the spreading gain. The cancellation order is that
user $K$ is detected first, then user $K-1$ and so on.
\end{proposition}

In the denominator in the argument of the $Q$-function in (\ref{BER_SC}), if errors exist for the
previously detected users, the interference caused to the latter detected
users is at  four times  the power level of the original signal. So if the
error probabilities of the previously detected users can be reduced by
cooperative transmission, the overall performance can be greatly
improved.

Notice that the BERs in (\ref{BER_Opt}) and (\ref{BER_SC}) are
functions of the users' received amplitudes. These in turn are
functions of the user locations and the network topology, which
are fixed in traditional multiuser channels. As will be shown
later, in our proposed schemes, we have the freedom to select
which users will serve as relays and which users' information to
relay. This freedom allows us to modify the link qualities and
achieve the optimal performance in terms of overall BER at the
base station.


%
%
%
%

\section{Two Cooperative Transmission Protocols\label{sec:Protocol}}

In this section, we propose two cooperative transmission
protocols. The first protocol seeks to exploit the fact that MUD
can improve the reception of all signals because of the mitigation
of interference from the strong ones. MUD is used in the base
station, while at the relay single user detection is employed. The
second protocol further exploits network coding in the relay to
make full use of the relay-destination channel and to provide
better coding gain and diversity gain. In this protocol, MUD is
employed at both the base station and the relay.

\subsection{Protocol 1: Joint MUD and Cooperative Transmission}

Suppose terminal $i$ is selected as the relay and it forwards user
$m$'s information. At the base station, following a matched filter
bank, maximal ratio combining (MRC) is used to combine the signals
from these two terminals. Since optimal MUD and decision driven
MUD algorithms are nonlinear, a closed-form expression for MRC is
not available. In our analysis, we assume that some method, such
as a threshold test \cite{bib:Laneman2} or cyclic redundancy check
(CRC), is employed so that the potential relays and the base
station can determine with some certainty whether or not the
detected signals are correct. Instead of MRC before decoding, the
final decision is based on the decoded signals in both stages.
Thus, an error occurs only if the signals in both stages are
wrong. So the probability of error can be written as
\begin{equation}\label{protocol_one_BER}
P_r^m=P_r^{m0}(1-(1-P_r^{mi})(1-P_r^{i0})).
\end{equation}
 The error probabilities of
data transmission from user $m$ to the base station, from user $i$ (i.e., the relay)
to the base station, and from user $m$ to user $i$ are denoted as
$P_r^{m0}$, $P_r^{i0}$, and $P_r^{mi}$, respectively. Notice that
there is no need for MUD at the relays for the first protocol.

\begin{table}
\caption{Cooperative Transmission Protocols} \label{solution}
\begin{center}
\begin{tabular}{|l|}
  \hline
  1. At Stage 1, the sources send packets, the base station stores \\
  \ \ \ them, and the relays decode them.\\
  \hline
  2. After Stage 1, the base station decides which users' packets \\
  \ \ \ are to be relayed according to (\ref{prob_def1}).\\
\hline
  3. At Stage 2, using the feedback from the base station, the \\
  \ \ \  relays forward the selected users' information so as to\\
  \ \ \  optimize the decoding. \\
\hline
  4. After Stage 2, two-stage combining and MUD are performed\\
  \ \ \  at the base station. \\
  \hline
\end{tabular}
\end{center}
\end{table}

The issues to be considered here are which relays to select among
the potential users (selecting $i$), and whose data to retransmit
(selecting $m$). The performance index for system optimization is
the overall BER. If only one relay is selected\footnote{For the
multiple relay case, if no user's information can be relayed more
than once, the problem formulation is the same. Otherwise, we need
to change (\ref{protocol_one_BER}). As a result, the searching
space will increase exponentially with the number of relays. In
that case, low complexity heuristics must be developed, which is
beyond the scope of this paper.}, the problem formulation to
minimize the overall BER can be written as
\begin{equation}\label{prob_def1}
\min _{(i,m)} \sum_{j\in \{ \mathfrak{K}\backslash i\}} P_r^j.
\end{equation}

\begin{figure}[htbp]
\begin{center}
    \includegraphics[width=80truemm]{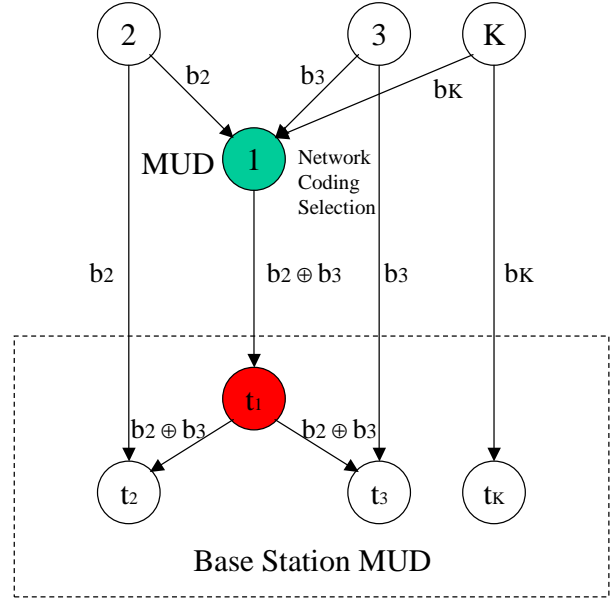}
\end{center}
\caption{Joint consideration of MUD and network
coding}\label{network_coding_model}
\end{figure}

To optimize (\ref{prob_def1}), we propose an algorithm shown in
Table \ref{solution}. The basic idea is that the base station can
know after Stage 1 which users' links need to be improved so as to
maximize the network performance. Moreover, the information of the
relay such as $P_r^{mi}$ and $P_r^{i0}$ can also be feeded back to
the base station. So the optimal parameter pair $(i,m)$ can be
selected\footnote{Here we use the exhaustive search for the
optimal pair. Some heuristic fast algorithms such as greedy
solution can be easily constructed.}, and the corresponding
information is sent. At the base station, the information sent at
the first stage is stored and combined with the relay's
information at the second stage. Consequently, the performance of
all users can be improved. The only control signaling required is
to send information through a control channel to inform the
corresponding relay which user's information to forward.

\subsection{Protocol 2: With Consideration of Network Coding}

The second protocol seeks to exploit the fact that MUD in the base
station and the relay provides a possible data-flow structure for
jointly optimizing MUD and network coding. In Figure
\ref{network_coding_model}, we illustrate an example in which
there are $K$ users and user $1$ is assigned as the relay. At the
first stage, users $2$ through $K$ send their own information,
while the base station and user $1$ listen. At the second stage,
user $1$ sends the coded information (here $b_2 \bigoplus b_3$,
where $\bigoplus$ is XOR function). Then the base station can
improve the decoding of user $2$ and user $3$. The performance
gain is due to the network coding.

In general, we can formulate joint MUD and network coding as
follows: As a relay, user $i$ selects a set of users
$\mathfrak{M}_i$, and then transmits $b_m \bigoplus \dots
\bigoplus b_n$, where $m,\dots, n\in \mathfrak{M}_i$. Notice that
$\mathfrak{M}_i$ is a subset of all users that are successfully
decoded at the first stage by user $i$. At the base station, the
user's error probability is given by:
\begin{eqnarray}\label{coded_user_m}
P_r^m=P_r^{m0}\{1-(1-P_r^{mi})(1-P_r^{i0}) \nonumber \\ \cdot
\prod _{n\in \mathfrak{M}_i/m}[(1-P_r^{ni})(1-P_r^{n0})]\},
\forall m\in \mathfrak{M}_i
\end{eqnarray}
and
\begin{equation}\label{other_user}
P_r^j\leq P_r^{j0},\ \forall j\notin \mathfrak{M}_i.
\end{equation}
The first term in (\ref{coded_user_m}) represents the direct
transmission error probability. The term in the parentheses of
(\ref{coded_user_m}) represents the error probability from the
relay using network coding. Successful transmission from
the relay occurs only if, without network coding gain, all
users in $\mathfrak{M}_i$ are decoded correctly by user $i$, the
transmission from user $i$ to the base station is correct, and all
other users are correctly decoded at the base station. Notice that
compared with (\ref{protocol_one_BER}), the error probability for
a specific user might be worse. However, since in
(\ref{coded_user_m}), multiple users' BERs can be improved, the
overall BER of the system can be further improved under careful
optimization. The inequality in (\ref{other_user}) holds since the
cancellation of some successfully decoded users' information can
improve the other users' decoding.

We need to select relay $i$ from the set $\mathfrak{R}$ of size
$N$, and the set $\mathfrak{M}_i$ which represent whose
information should be relayed by user $i$. So the general problem
formulation for both Protocol 1 and Protocol 2 can be written as
\begin{equation}\label{prob_def}
\min _{\mathfrak{R},\mathfrak{M}_i} \sum_{j\in \{
\mathfrak{K}\backslash \mathfrak{R}\}} P_r^j.
\end{equation}

The algorithm for Protocol 2 is similar to that of Protocol 1
except that, in Protocol 2, the relays transmit the network coded
symbol in the second stage. Compared with the first protocol,
Protocol 2 can improve more than one users' signal strength at the
base station in Stage 2. This is because several users'
information can be carried using network coding. However, if
 too many users' information is coded with network coding, the
error correction capability in the base station will be reduced.
So there is a tradeoff on how many users' information to be
encoded. Moreover, Protocol 2 requires MUD at the relay which could be
a mobile handset. Since this requirement increases the cost
and power consumptions of relays, this could be an issue.

\section{Performance Analysis\label{sec:analysis}}

In this section, we first examine the diversity order and coding
gain of the proposed protocols. Then, we give a performance upper
bound using MIMO-MUD. Next, we study a special case for how the
relay changes the asymptotic multiuser efficiency. Finally, we give an
exact expression for a symmetric case.

\subsection{Diversity Order and Cooperative MUD Gain}

First we study the diversity order for the users whose information
is relayed. Then we provide another performance gain metric,
cooperative MUD gain, to quantify the additional gain to the other
users.

For Protocol 1, the diversity order (i.e., the number of
independently received signals) can be up to $N+1$ for the relayed
user, while the remaining un-relayed users have diversity order
$1$. For Protocol 2, the diversity order for all users is up to
$N+1$. From the simulation results presented below, we see that
the high diversity order can be achieved compared to a performance
bound (which has been shown to have the high diversity order).
Rigorous proof for the high diversity order of the proposed scheme
is very difficult to achieve, due to the intractability of BER
expressions for MUD detectors. However, we provide an intuitive
analysis in the sequel.

For Protocol 1 after Stage 1, we order the received signals at the
base station according to their SINRs, where user $K$ has the
highest SINR (i.e. the lowest BER). We assume all $N$ relays
select user $K$'s information to retransmit if the relay decodes
it correctly. The reason to select user $K$ with the highest SINR
is to limit error propagation in (\ref{BER_SC}). The diversity
order for user $K$ is $N+1$ since $N+1$ copies of user $K$'s
information are transmitted via $1$ direct link and $N$ relay
links and all those link responses are independent. Because only
user $K$'s copy of the information at Stage 1 is retransmitted,
the diversity order of the other users is still $1$. If the $N$
relays select different users' information to relay, the diversity
orders of these users depend on how many relays retransmit their
information.

For Protocol 2, at the second stage the relays retransmit the
following information
\begin{equation}
z_i=\bigoplus b_j, j\in \mathfrak{M}_i.
\end{equation}
Here we assume $\mathfrak{M}_i$ includes all users, i.e.,
$\mathfrak{M}_i=\mathfrak{K}$.

When the SINRs are sufficiently high (i.e. the multiple access
interference is sufficiently low), the channels between the
senders and relays approach ideal links. All direct links are
independent and approach ideal links. For example in Figure
\ref{network_coding_model}, at the second stage after network
decoding, $t_2$ will receive two copies of $b_2$ from direct
transmission and from $t_1$ if $b_3$ has sufficiently small BER.
In a generalized case, if the size of $\mathfrak{M}_i$ is $K$, the
diversity order for every user is $N+1$ from the $N$ relays and
the direct link, when the SINRs are sufficiently high. Another
interpretation is that when the SINRs become sufficiently large,
the links between the relays and base station are sufficiently
good. Consequently, the cooperative system with Protocol 2 is
equivalent to MIMO MUD system with diversity order of $N+1$.

On the other hand, if the diversity orders of certain users
increase, the remaining users have better performance since their
interference (user $K$'s signal) can be more successfully
cancelled. To quantify the performance gain, we define the
following quantity.
\begin{definition}
{\em The cooperative MUD gain} $\rho_i$ is defined as the SINR
improvement ratio for the remaining users, due to the link
improvement gained when the other users use cooperative MUD
receivers.
\end{definition}

For the successive cancellation multiuser detector of Protocol 1, we
have
\begin{equation}\label{MUD_gain_SC}
\rho_{K-1}=\frac{\sigma^2+ \frac 1 M\sum_{j=1}^{K-2} A_j^2+ \frac
4 M A_{K}^2 P_r^K}{\sigma^2+ \frac 1 M\sum_{j=1}^{K-2} A_j^2+
\frac 4 M A_{K}^2 \hat P_r^K},
\end{equation}
where $\hat P_r^K$ is user $K$'s new BER and $\hat P_r^K\approx
(P_r^K)^{N+1}$. If $\frac 1 M A_K^2 >> \sigma^2+ \frac 1 M
\sum_{j=1}^{K-2} A_j^2$, the MUD gain can be significantly large.
For the MUD gains of other users, we can calculate
$P_r^{K-1},\dots, P_r^1$ recursively.

For the optimal MUD  of Protocol 1, for each possible
error vector $\epsilon$, the MUD gain can be approximated by
\begin{equation}
\rho^{\epsilon}\approx \epsilon ^T \mbox{diag} \{A_i,\hat
A_j\}\textbf R \mbox{diag} \{A_i,\hat A_j\} \epsilon,
\end{equation}
where $\hat A_j$ is the improvement of the $j^{th}$ user's signal
strength and $\mbox{diag} \{A_i,\hat A_j\}$ is the same as matrix
$\textbf A$ (defined in (\ref{A})) except that $A_j$ is replaced
by $\hat A_j$. Notice that the channel improvement $\hat A_j$  is
upper bounded by that of MRC of direct transmission and relay
transmission.

For the successive cancellation detector of Protocol 2, the MUD
gain for the user with the second strongest link is
the same as (\ref{MUD_gain_SC}). For the remaining users, the MUD
gain is larger since higher diversity order for all the users with
larger SINR reduces the error probabilities, which affect the noise
of this user. For the optimal MUD  of Protocol 2, the
elements of the matrix $\textbf A$ increase (i.e.,  every link
$A_i$ is enhanced with diversity $N+1$). So the BER for each
possible error vector is also reduced and so is the overall BER.

\subsection{Performance Bounds}

We develop two performance bounds for the proposed cooperative
transmission protocol with MUD. First, in MIMO MUD
\cite{HuaiyuDai}, we can assume infinite bandwidth between the relays
and the base station. The performance under these circumstances
gives us an upper bound for
Protocol 2 of cooperative transmission MUD. Here we assume that
the relay is perfectly connected to the destination,  and that
combination is performed after decoding. Decoding error occurs
when the direct transmission and all of the $N$ source-relay links
fail, i.e. for the  high SINR case we have the first performance upper
bound given by
\begin{equation}\label{bound1}
P_r^k\geq  P_r^{k0} \prod _{i\in \mathfrak{R}} P_r^{ki},
\end{equation}
where $P_r^{k0}$ is the BER for direct transmission and
$P_r^{ki}$ is the transmission from user $k$ to relay $i$. For
MIMO MUD, the diversity order is $N+1$.

Second, if we assume that the links between the source and relays
are perfect and the SINRs for the two stages can be directly
added before the decoding, we can obtain  another performance
upper bound for Protocol 1. If we assume all relays retransmit
user $k$'s information, for the successive cancellation detector, we can derive a bound in the following recursive form:
\begin{equation}\label{bound2}
P_r^{k,sc}\geq  Q \left( \frac {A_k+\sum _{i\in \mathfrak{R}}
A_i}{\sqrt{\sigma^2+\frac 1 M \sum_{j=1}^{i-1}A_j^2+\frac 4 M
\sum_{j=i+1}^K A_j^2P_r^{j,sc}}}\right) .
\end{equation}
Notice that the interference terms from the stronger users in the
denominator still have amplitudes $A_j$'s, since the interference
comes from the first stage of the cooperative transmission. For
optimal MUD, the second performance upper bound can be
obtained by setting $\textbf A=\mbox{diag}\{ A_1, \dots, A_k+\sum
_{i\in \mathfrak{R}} A_i, \dots A_N\}$.

Another interpretation of the above two bounds is as follows. For
the bound in (\ref{bound1}), all relays are located close to the
base station so that the relay-destination links are sufficiently
good. For the bound in (\ref{bound2}), all source users and relays
are assumed to be closely located in a cluster away from the base
station. The source-relay links are assumed to be perfect. In
reality, the source nodes and relay nodes are located randomly. So
the real performance is worse than the two performance upper
bounds - i.e., the bounds may not be tight.

\subsection{Asymptotic Multiuser Efficiency}

In this subsection, we study a special case in which there are two users and one
relay to investigate the performance improvement that results from using MUD. First, we
review the definition of asymptotic multiuser efficiency.
\begin{definition}
The asymptotic multiuser efficiency is defined as
\begin{equation}
\eta _k=\lim _{\sigma \rightarrow 0} \frac {\sigma ^2}{A_{k}^2}
\log \left(\frac 1 {P_r^k}\right)
\end{equation}
which quantifies the degradation in SINR suffered by a user due to the
presence of other users in the channel.
\end{definition}

Similarly to the second performance upper bound in the previous
subsection, we make the approximations that the relay can always
decode correctly and  that the base station can use maximal ratio
combining of the direct and relay transmissions. In this ideal
case, the multiuser efficiency of optimal MUD  can been
expressed as
\begin{eqnarray}
\eta _1\approx \min \left\{ 1, 1+\frac {(A_2+A_r)^2}{A_1^2}-2|\rho
| \frac {A_2+A_r}{A_1},\right.\nonumber \\
\left. 1+\frac
{A_2^2}{(A_1+A_r)^2}-2|\rho | \frac {A_2}{A_1+A_r}\right\} ,
\end{eqnarray}
where $\rho$ is the cross-correlation, and $A_1$, $A_2$, and $A_r$
are the channel gains to the base station for user $1$, user $2$
and the relay, respectively.
%
%
%

\begin{figure}[htbp]
\begin{center}
    \includegraphics[width=80truemm]{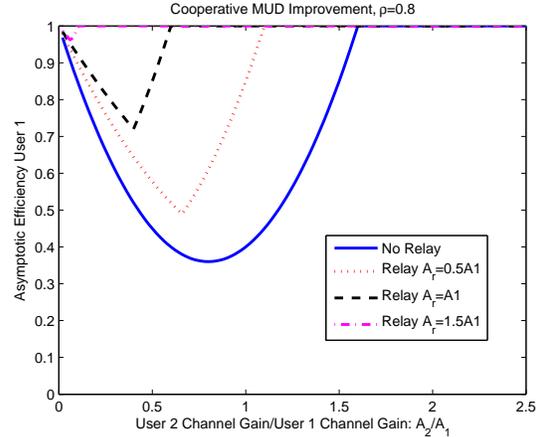}
\end{center}
\caption{Asymptotic multiuser efficiency improvement as a function
of users' channel gains $A_2/A_1$}\label{MUD_efficiency}
\end{figure}

In Figure \ref{MUD_efficiency}, we show the asymptotic multiuser efficiency with
$A_1$=1 and $\rho=0.8$. The key idea here is that the asymptotic
multiuser efficiency is bad when the ratio of $A_2$ and $A_1$ is
around $\rho$, but with the relay's help this ratio can be changed
so that the asymptotic multiuser efficiency can be greatly
improved. We can see that when the relay is close to the
destination (i.e. $A_r$ is large), the asymptotic multiuser
efficiency can be almost $1$. This is because the relay can always
improve the stronger user's link so that the difference is even
larger. Consequently, the multiuser efficiency can be greatly improved.
When the relay moves far away from the destination, the asymptotic
multiuser efficiency improvement is reduced, since $A_r$ decreases
and the relay is less effective. We note that this comparison is
unfair, since the bandwidth is increased with the presence of the
relay. However, when the number of users is sufficiently larger
than the number of relays, this increase is negligible.

\subsection{Special Case Analysis}

In this subsection, we study a special case to examine issues
 such
as how many relays should be used for network coding and which
relays should be selected. We consider the case in which several
source nodes are located close to each other and far away from the
base station. In this situation, the links between the sources to
one relay are the same and the links from the different sources to
the destination are equal. This special case fits the scenario in
which there is no base station in a community. The error
probabilities incurred in transmission from source to destination, from
 source to relay and from  relay to destination are
$P_r^{sd}$, $P_r^{sr}$, and $P_r^{rd}$, respectively. We assume
Protocol 2 is used and we suppose the relay includes $M$ out of
$K$ sources for networking coding. The coded users' error
probability is given by
\begin{equation}
P_r=P_r^{sd}[1-(1-P_r^{sr})^M(1-P_r^{sd})^{M-1}(1-P_r^{rd})].
\end{equation}
There are $K-M$ users without network coding gain and $M$ users
with network coding gain. To minimize the overall average BER, we
have
\begin{eqnarray}
\min _{M} \left\{ P_r^{ave}=\frac 1 K \{
P_r^{sd}(K-M)+MP_r^{sd}[1-\right.\nonumber\\
\left.(1-P_r^{sr})^M(1-P_r^{sd})^{M-1}(1-P_r^{rd})]\} \right\}.
\end{eqnarray}
It is easy to show that the optimal number of users to be included
in network coding is
\begin{eqnarray}\label{optimal_M}
M^*=\min\left\{K, \arg \min [ P_r^{ave}(M_1), P_r^{ave}(M_2)]\right.\nonumber\\
\left.\mbox{ where } M_1\leq \frac
{-1}{\log[(1-P_r^{sr})(1-P_r^{sd})]}\leq M_2 \right\}
\end{eqnarray}
where $M_1$ and $M_2$ are the two non-negative integers closest to
$\frac {-1}{\log[(1-P_r^{sr})(1-P_r^{sd})]}$.

From (\ref{optimal_M}), we can make the following observations.
First, if the source-to-relay and relay-to-destination channels
are relatively good, it is optimal to include all users in
network coding. For example, when $(1-P_r^{sr})(1-P_r^{sd})=0.99$,
as long as $K < 100$, it is optimal. Second, in order to minimize
$P_r$, the relay needs to have a large value of
$(1-P_r^{sr})(1-P_r^{sd})$. This fact suggests a relay selection
criterion in practice.

%
%
%


%
%

\subsection{Implementation Discussion}

In this subsection, we discuss some implementation issues. First,
our proposed protocols do not work for certain types of MUD. For
the decorrelating detector, the proposed schemes are not suitable,
since the performance is controlled by the cross correlation. For
the minimum mean square error (MMSE) receiver, the proposed scheme
is not effective, since the improvement of one user's detection
does not improve that of the others for linear detectors. MMSE
detector performance under cooperative communication is
investigated in \cite{Vojcic}. A variety of other MUD receivers
can still be used, such as the decision feedback MUD, multiple
stage MUD, blind MUD, and their combinations with the linear MUD.
But, this complicates the analysis of the proposed schemes due to
the nonlinearity of these other MUD techniques.

Second, we discuss the asymptotic behavior of large systems
\cite{Asym1}\cite{Asym2}\cite{Asym3}. Denote by $\beta$  the
system load  (i.e., the number of active users divided by the
number of codes in CDMA). For  the decorrelator, the multiuser
efficiency is given by
\begin{equation}
\mbox{decorrelator: }\eta=1-\beta.
\end{equation}
We can see that our proposed scheme cannot work at all in this
case. For the MMSE detector, the multiuser efficiency is obtained by
solving the following equation:
\begin{equation}\label{MMSE_asym}
\mbox{MMSE MUD: }\eta+\beta E\left( \frac{\eta P}{\sigma ^2_n
+\eta P}\right)=1,
\end{equation}
where $\sigma ^2_n$ is the noise power level and $P$ is the
received power, over which the expectation is carried. From
(\ref{MMSE_asym}), we can see that our proposed scheme can improve
the relayed users' random received powers. So the resulting $\eta$
is larger. However, this improvement is for the relayed users only
and cannot ``propagate" to benefit the other users. For
optimal MUD, the following equations
\cite{Asym1}\cite{Asym2}\cite{Asym3} can be solved with the
variables are $E$, $F$, $m$, and $q$:
\begin{eqnarray}
E&=&\frac{1}{\sigma^2+\beta(1-m)},\label{optimal_asym1}\\
m&=&1-E\left(\frac P {1+PE}\right),\\
F&=&\frac{\sigma^2_n+\beta(1-2m+q)}{[\sigma ^2+\beta(1-m)]^2},\\
q&=&E\left[
\frac{P^3E^2+P^2F}{(1+PE)^2}\right],\label{optimal_asym4}
\end{eqnarray}
where $\sigma^2$ equals $\sigma ^2_n$ when individual MUD is
used and equals 0 when joint MUD is used. Then, the multiuser
efficiency is obtain by
\begin{equation}\label{optimal_asym}
\mbox{optimal MUD: }\eta=\frac{E^2}{F}\sigma ^2 _n.
\end{equation}
From (\ref{optimal_asym1}) to (\ref{optimal_asym4}), we can see
that if the expectation over the random received power is improved
by the proposed scheme, the parameters affect each other
recursively. As a result, the multiuser efficiency in
(\ref{optimal_asym}) can be greatly improved. This is another
demonstration of our main idea that improving one user's link can
benefit the others.

Finally, we discuss some practical implementation issues and how
the proposed schemes can be integrated into existing networks such
as cellular networks. Because of handware limitations, it is often
difficult to implement  MUD   in a mobile terminal. However, we can
implement Protocol 1 in the mobile terminal to relay the other
users' information. In the base station, the MUD
performance can thereby be improved. To optimally improve the system
performance, the issues of relay selection and whose information
to relay need to be solved. If a service provider can set up fixed
relays which are much cheaper than the base station, the second
protocol can be employed to have MUD in the fixed relays.
Moreover, network coding can be used to provide full diversity gain.
The issues of where the fixed relays should be located and how
many users should participate in network coding need to be examined. The
simulations in the next section examine all of these issues.

\section{Simulation Results}\label{sec:simulation}

In order to demonstrate the effectiveness of the proposed
protocols, we present simulations with the following setup. First,
we consider a one-dimensional model in which a base station, a
relay, and users are located along a line. The base station is
located at position $0$ in the coordinate system, the two users
are located at position $4$ and position $6$, and the relay can
move from position $0.5$ to position $3.5$. The loss factor for
large scale propagation is $3$. In the simulation, we assume that
all users and the relay use the same transmitted power, i.e.,
there is no power control. We also assume the receivers have the
same additive noise with power level $0$dB. MUD is used only for
Protocol $2$.


\begin{figure*}
\footnotesize
        \subfigure[Successive cancellation] {
        \label{fig:mean_Pe_vs_Tpower_relay1p6}
        \includegraphics[width=80truemm]{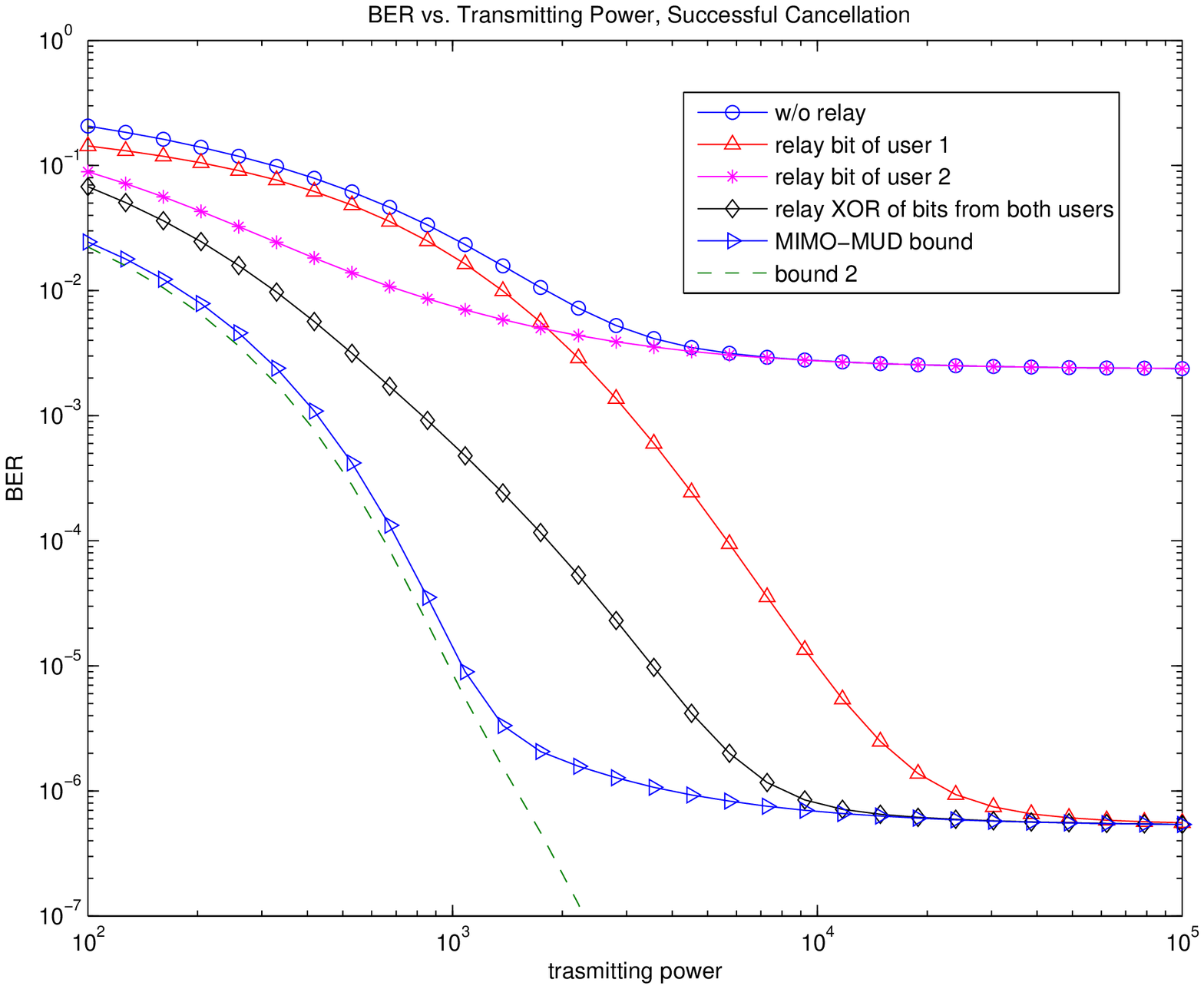}}
        \subfigure[Optimal MUD] {\label{fig:mean_Pe_vs_Tpower_relay1p6_opt}
        \includegraphics[width=80truemm]{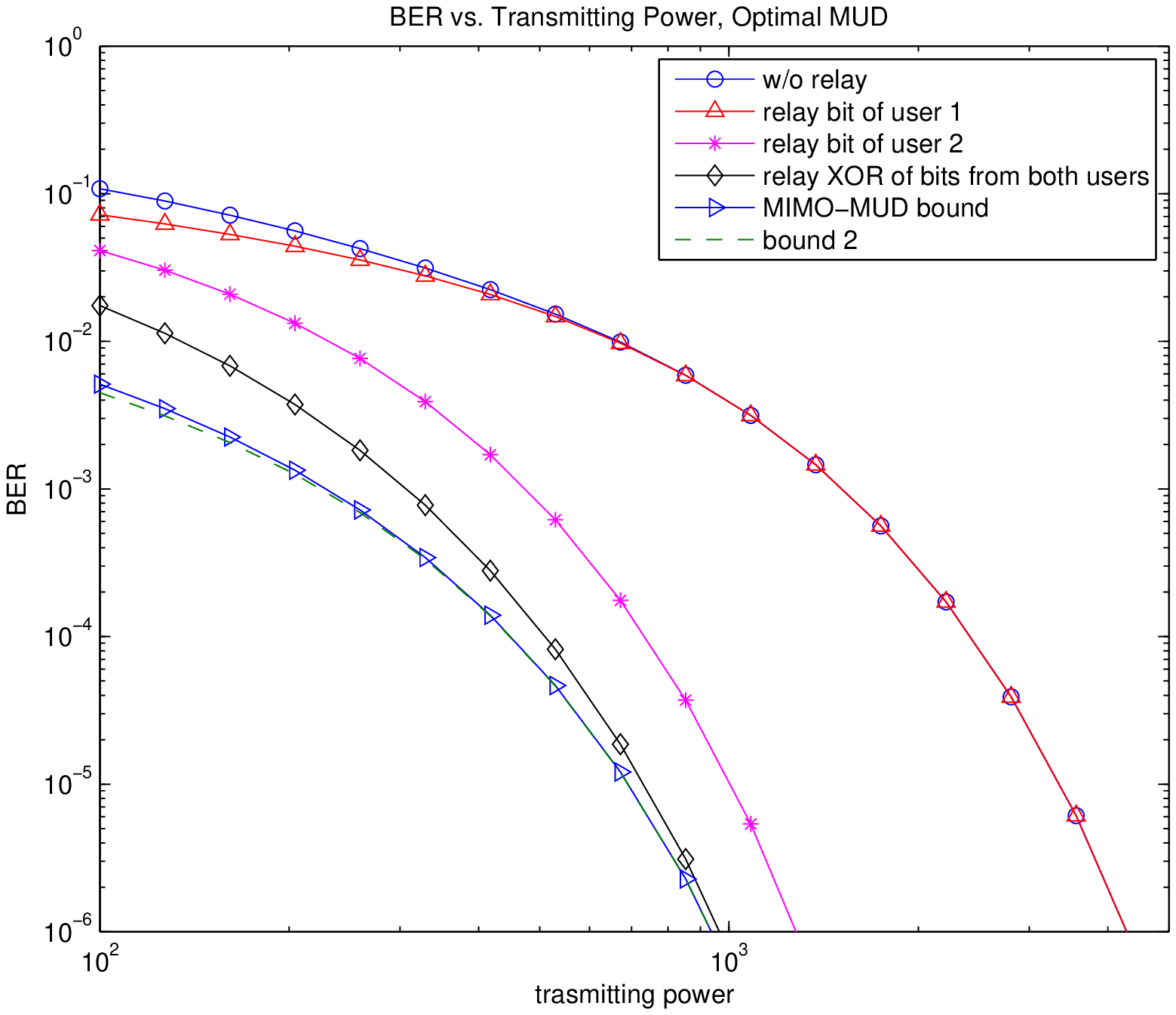}}
            \caption{The average BER as a function of the transmitted power with the relay located at 1.6.}
\end{figure*}

Figure \ref{fig:mean_Pe_vs_Tpower_relay1p6} and Figure
\ref{fig:mean_Pe_vs_Tpower_relay1p6_opt} show the average BER at
the base station as a function of the transmitted power of the
users and of the relay for successive cancellation  and optimal
MUD, respectively. The relay's location is fixed at position
$1.6$. We can clearly see the higher diversity order of BER vs.
power for the proposed protocols. When the transmitted power is
sufficiently high, the limiting factor of  successive
cancellation's performance is the interference. And that is why we
see the curve flattens when the transmitted power grows. We also
note the large difference in performance between the case with the
relay and the case without. Another interesting observation is
that, for successive cancellation  in a certain transmitted power
range, relaying the first user's symbol is better, while in
another transmitted power range, relaying the second user's symbol
is better. For optimal MUD, to relay the symbol of user $2$ is
always the best choice. Relaying the XOR of both users' symbols is
always the best protocol, but this requires the use of MUD at the
relays. We also show the MIMO-MUD performance bound and bound $2$
which assumes perfect channels from source to relay. The two
bounds are similar except when successive cancellation  hits an
error floor. The bounds for optimal MUD are  tighter especially
when the BER is sufficiently low. When $\mbox{BER}=10^{-6}$, for
optimal MUD , the performance gap between the bounds and the
protocol in which the relay XORs  bits from both users is
$0.12$dB. When $\mbox{BER}=10^{-3}$, the gap is $1.04$dB.

\begin{figure*}
\footnotesize
        \subfigure[Successive cancellation, power $= 40$dB] {
        \label{fig:mean_Pe_vs_relaylocation_high_SNR}
        \includegraphics[width=80truemm]{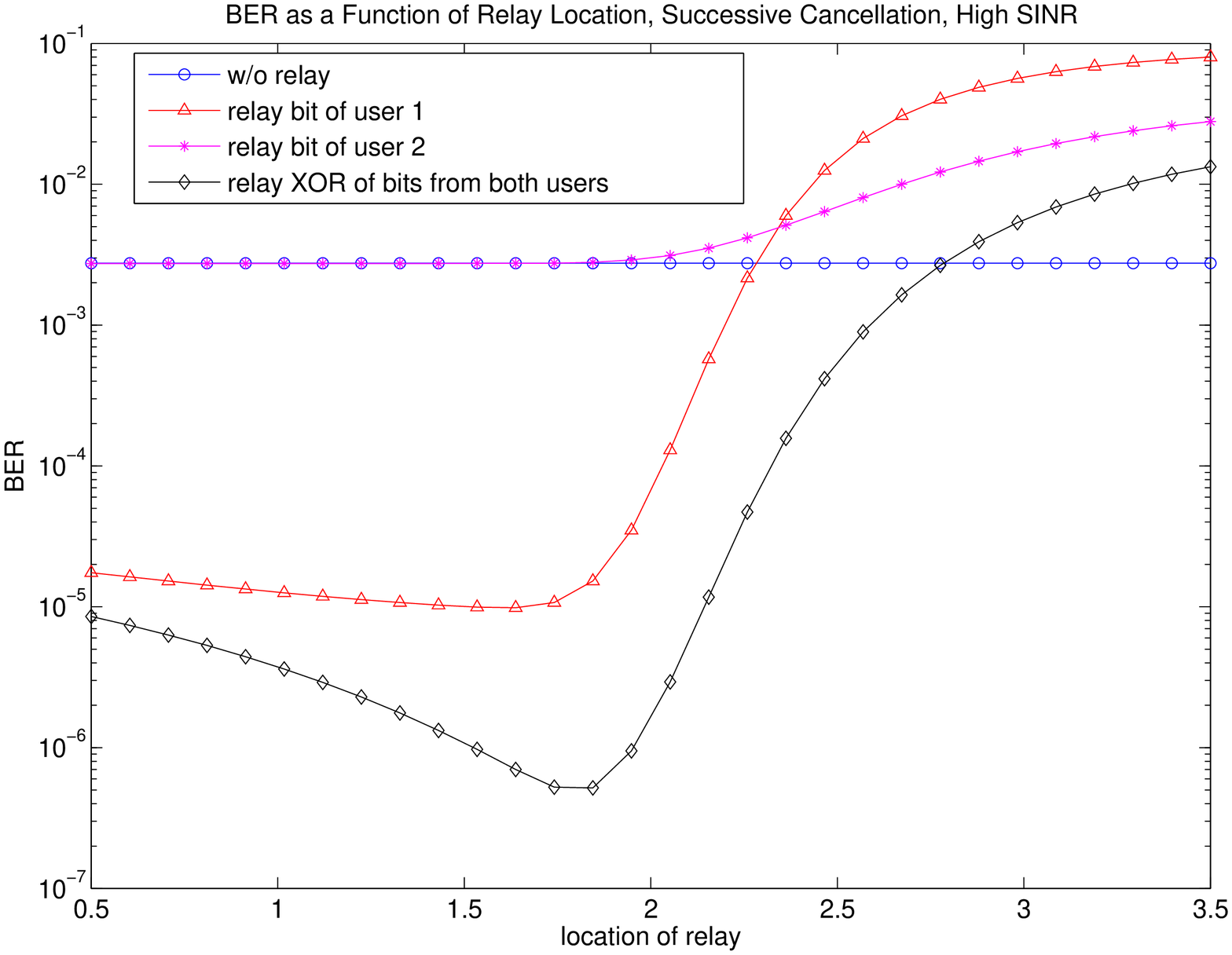}}
        \subfigure[Optimal MUD, power $= 30$dB] {\label{fig:mean_Pe_vs_relaylocation_high_SNR_opt}
        \includegraphics[width=80truemm]{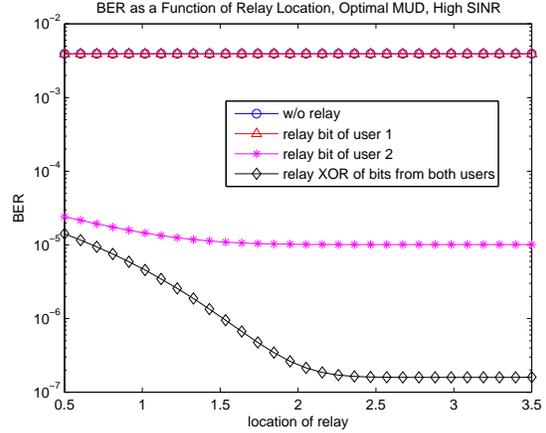}}
            \caption{The average BER as a function of the location of the
relay with high average SINR.}
\end{figure*}

Figure \ref{fig:mean_Pe_vs_relaylocation_high_SNR} and Figure
\ref{fig:mean_Pe_vs_relaylocation_high_SNR_opt} show the average
BER at the base station as a function of the relay location. There
are two users, and both users and the relay use high transmitted
power of $40$dB for successive cancellation  and $30$dB for
optimal MUD. The curves correspond to the case without the relay,
with the relay re-transmitting user 1's (located at position $4$)
symbol, with relay re-transmitting user 2's (located at position
$6$) symbol, and with the relay re-transmitting the XOR of both
users' symbols (network coding), respectively.

The first observation is that the location of the relay plays a
vital role in the system performance, especially for the
successive cancellation detector. For successive cancellation,
 the system with a relay performs better than the system
without a relay, only if the relay's distance from the base
station is below position $2.8$ using network coding and below
position $2.2$ when the relay helps user 1. If the successive
cancellation detector is used, the system performs better without
a relay if the relay is too close to the user group. This is
because, for successive cancellation, the performance is better if
the users have different received power levels. A relay that is
too close to the user group will increase the error rate of the
successive cancellation detector because of its interference. On
the other hand, for optimal MUD, the performance is always better
with a relay, especially when the relay is close to the users.
However, the performance improvement has a floor. The second
observation is that there is a ``sweet spot'' for successive
cancellation eith the location of the relay around position 1.8.
This is because the relay's decoding performance drops if it is
located too far away from the sources. The third observation is
that the network coding protocol with the relay re-transmitting
the XOR of both users' symbols always performs better than that
when the relay just re-transmits one user's symbol.

\begin{figure*}
 \footnotesize\centering
        \subfigure[Successive cancellation, power $= 30$dB] {
        \label{fig:mean_Pe_vs_relaylocation_low_SNR}
        \includegraphics[width=80truemm]{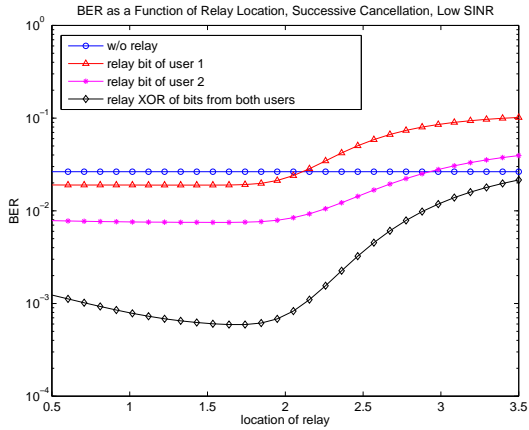}}
        \subfigure[Optimal MUD, power $= 20$dB] {\label{fig:mean_Pe_vs_relaylocation_low_SNR_opt}
        \includegraphics[width=80truemm]{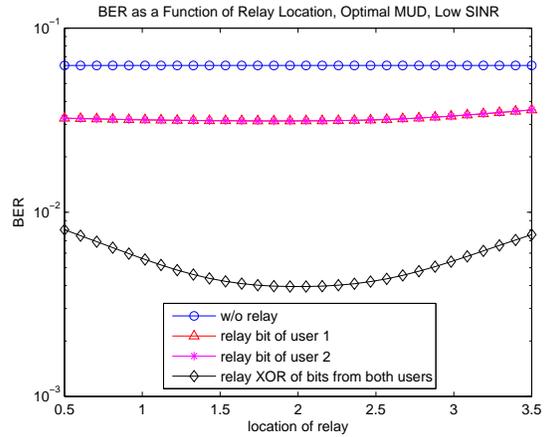}}
            \caption{The average BER as a function of the location of the
relay with low average SINR}
\end{figure*}

Figure \ref{fig:mean_Pe_vs_relaylocation_low_SNR} and Figure
\ref{fig:mean_Pe_vs_relaylocation_low_SNR_opt} correspond to
similar setups except the transmitted power is low here ($30$dB
for successive cancellation and $20$dB for optimal MUD). For
successive cancellation, we observe performance behavior similar
to the high transmitted power case, except that the relay can
still help when its location is close to the users. The ``sweet
spot" remains essentially at the same place. For optimal MUD,
there exists a ``sweet spot" as well at the position around $2$.
From the network designer's point of view, if a fixed relay can be
added to the network to improve the performance, the above
observations on the relay locations can provide  guidance on where
to place such a fixed relay.

\begin{figure*}
 \footnotesize\centering
        \subfigure[Successive cancellation] {
        \label{fig:mean_Pe_vs_num_user}
        \includegraphics[width=80truemm]{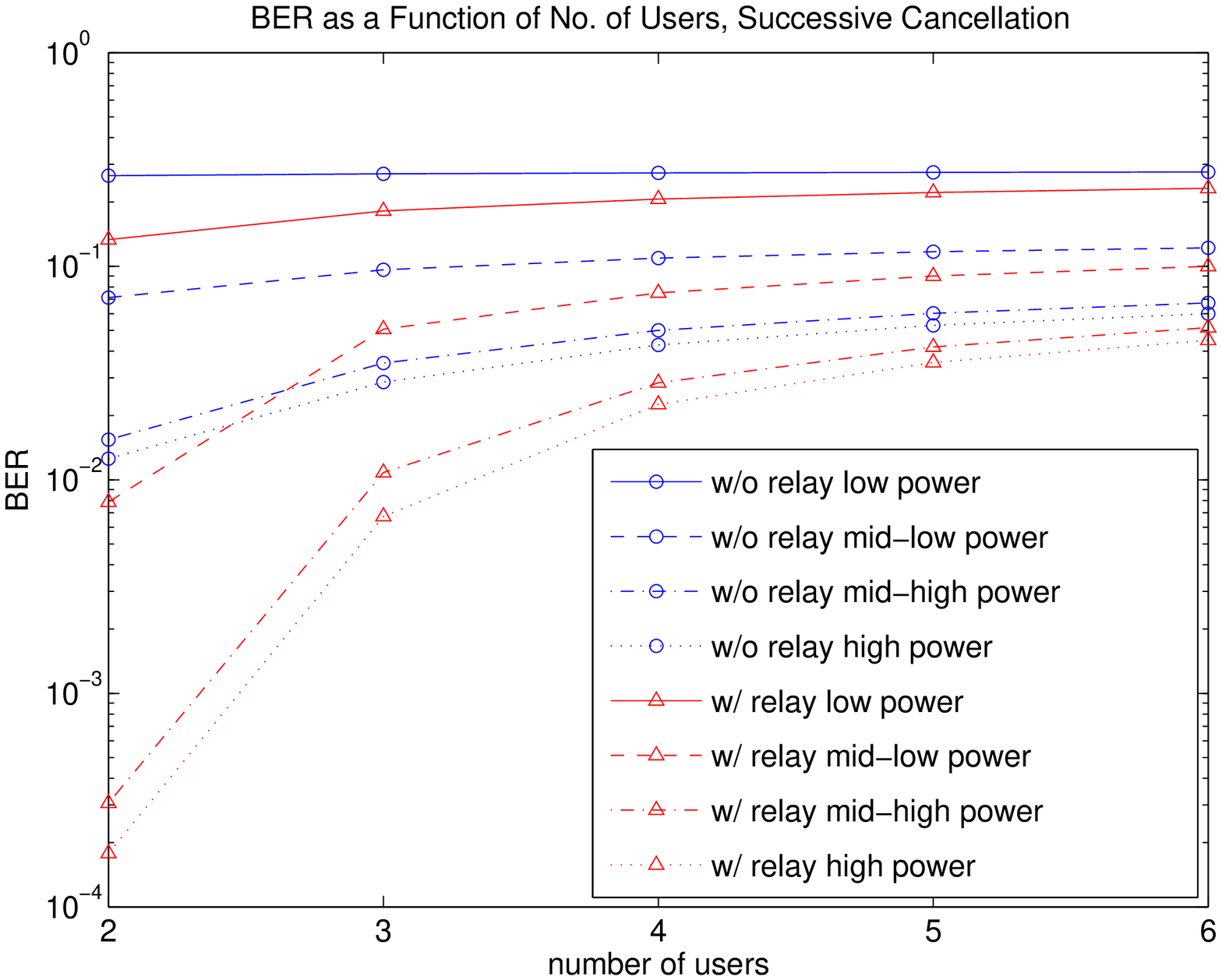}}
        \subfigure[Optimal MUD] {\label{fig:mean_Pe_vs_num_user_opt}
        \includegraphics[width=80truemm]{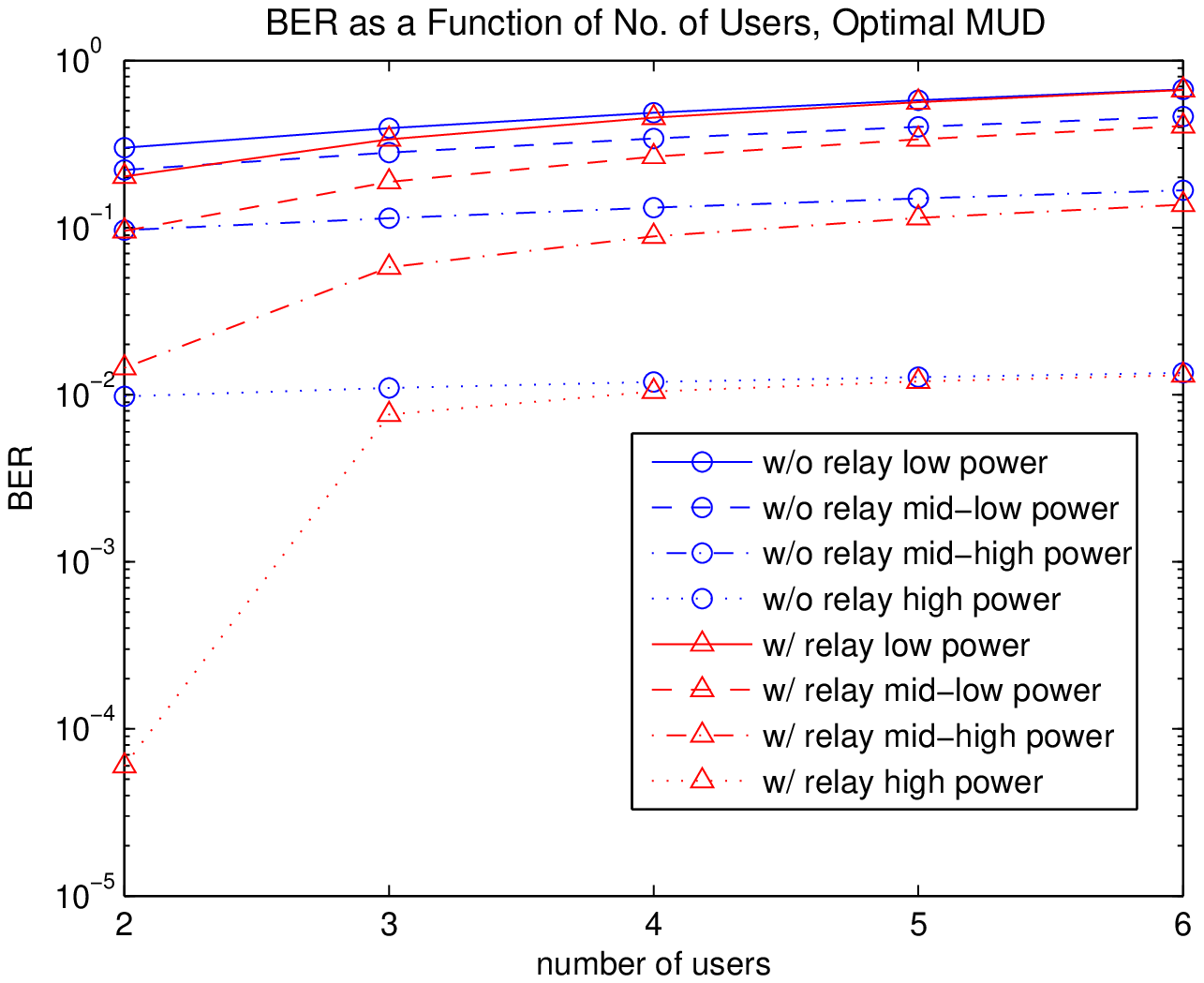}}
            \caption{The average BER as a function of the number of users uniformly
distributed in the range $[4,8]$ and with the relay  located at
$1.6$. The power settings are $20$dB, $30$dB, $40$dB, and $50$dB
for the lower, mid-low, mid-high, and high power setups,
respectively.}
\end{figure*}

Figure \ref{fig:mean_Pe_vs_num_user} and Figure
\ref{fig:mean_Pe_vs_num_user_opt} show the average BER as a
function of the number of users. Here we explore the cases with
two to six users. In each case, the users are uniformly
distributed in the range $[4,8]$. The relay is located at $1.6$,
and transmits the XOR of the nearest two users' symbols. For
successive cancellation, the power settings are $20$dB, $30$dB,
$40$dB, and $50$dB for the lower, mid-low, mid-high, and high
power setups. For optimal MUD, the power settings are $10$dB,
$16.7$dB, $23.3$dB, and $30$dB instead. As expected, the
performance is best when there are only two users. The performance
for the case with more users can be improved by introducing more
relays or having the relay transmitting XOR of more users'
symbols.

\begin{figure}[htbp]
\begin{center}
    \includegraphics[width=80truemm]{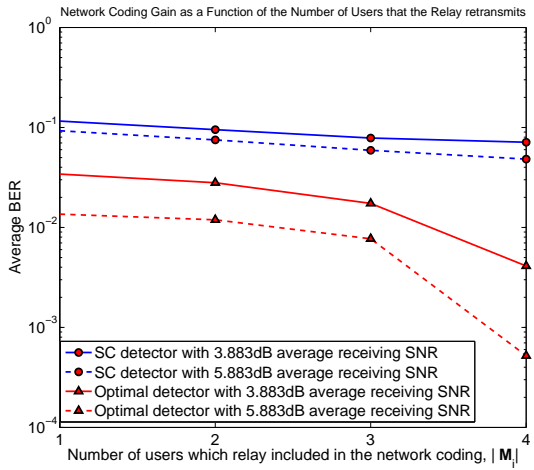}
\end{center}
\caption{Average BER as a function of the number of users for the
relay to be coded with network coding with different average SNR
and different MUDs} \label{fig:BER_vs_num_user}
\end{figure}

Finally, we study the problem of whose information should be coded
with network coding. Figure \ref{fig:BER_vs_num_user} shows the
average BER as a function of the number of users for the relay to
be coded with network coding with different average SNR and
different MUDs. We can observe that, in this case, coding
more users can improve the system performance. From the above
observations, we can see that the system performance degrades as
the number of users in the network increases, while the proposed
approach with network coding and cooperative MUD can significantly
improve the performance by encoding more users.

\section{Conclusions}\label{sec:conclusion}

In this paper, based on the fact that the enhancement of some
users' transmissions by cooperative transmission can improve the
other users' performance in certain types of multiuser detectors,
we have proposed two new cooperative transmission protocols that
utilize MUD as well as network coding. Unlike traditional MUD in
which the links are determined by the users' locations and
channels, the proposed cooperative transmission protocols improve
the link qualities so that the multiuser detectors can work in
their most efficient regions. Moreover, deploying MUD at the relay
provides an opportunity to use network coding, which can provide
additional coding gain and achieve full diversity. From our
analytical and simulation results, it is seen that the proposed
protocols achieve much lower average BER, higher diversity order
and coding gain, and better asymptotic efficiency, compared to
cooperative transmission in networks using single user detection
and traditional MUD. The performance gap between the proposed
approach and the MIMO-MUD bound is less than $0.12$dB when the BER
is $10^{-6}$ and 1.04dB when the BER is $10^{-3}$.

\section*{Acknowledgment}

The authors would like to thank Dr. Huaiyu Dai of North Carolina
State University, Dr. Husheng Li of  the University of Tennessee,
and anonymous reviewers for their constructive discussions and
suggestions.

\bibliographystyle{IEEE}

\begin{thebibliography}{1}

\bibitem{bib:Aazhang1}
A. Sendonaris, E. Erkip, and B. Aazhang, ``User cooperation
diversity, Part I: System description,'' {\it IEEE Transactions on
Communications}, vol. 51, no. 11, pp. 1927-1938, November 2003.

\bibitem{bib:Laneman2}
J. N. Laneman, D. N. C. Tse, and G. W. Wornell, ``Cooperative
diversity in wireless networks: efficient protocols and outage
behavior,'' {\it IEEE Transactions on Information Theory}, vol.
50, no. 12, pp. 3062-3080, December 2004.

\bibitem{bib:ICC06} A. K. Sadek, Z. Han, and K. J. Ray Liu, ``Distributed
relay assignment algorithm for cooperative communications in
wireless networks,'' in {\em Proc. IEEE International Conference on
Communications}, Istanbul, Turkey, June 2006.


\bibitem{Luo} J. Luo, R. S. Blum, L. J. Greenstein, L. J. Cimini, and A. M. Haimovich,
``New approaches for cooperative use of multiple antennas in ad
hoc wireless networks," in {\em Proc. IEEE Vehicular
Technology Conference}, vol.4, pp.2769- 2773, Los Angeles, CA,
September 2004.

\bibitem{bib:WCNC06_sensor} T. Himsoon, W. Siriwongpairat, Z. Han, and K. J.
Ray Liu, ``Lifetime maximization with cooperative diversity in
wireless sensor networks," in {\em Proc. IEEE Wireless Communications
and Networking Conference}, Las Vegas, NV,  April 2006.

\bibitem{bib:Madsen}
Z. Yang, J. Liu, and A. Host-Madsen, ``Cooperative routing and
power allocation in ad-hoc networks,'' in {\it Proc. IEEE
Global Telecommunications Conference}, Dallas, TX, November 2005.

\bibitem{bib:WCNC06_WIFI} Z. Han, T. Himsoon, W. Siriwongpairat, and K. J. Ray
Liu, ``Energy efficient cooperative transmission over multiuser
OFDM networks: who helps whom and how to cooperate," in {\em
Proc. IEEE Wireless Communications and Networking
Conference}, vol. 2, pp.1030-1035, New Orleans, March 2005.

\bibitem{bib:WCNC06_UWB} W. Siriwongpairat, W. Su, Z. Han, and K. J. R. Liu,
``Enhancement for multiband UWB systems using cooperative
communications," in {\em Proc. IEEE Wireless Communications and
Networking Conference}, Las Vegas, NV,  April 2006.





%
%

\bibitem{verdu} S. Verd\'{u}, {\em Multiuser Detection}, Cambridge
University Press, Cambridge, UK, 1998.

\bibitem{Venturino} L. Venturino, X. Wang and M. Lops, ``Multiuser detection for
cooperative networks and performance analysis," {\em IEEE
Transactions on Signal Processing}, vol. 54, no. 9, pp. 3315-3329,
September 2006.

\bibitem{HuaiyuDai}   H. Dai, A. F. Molisch, and H. V. Poor,
``Downlink capacity of interference-limited MIMO systems with
joint detection," {\em IEEE Transactions on Wireless Communications},
vol. 3, no. 2, pp. 442-453, March 2004.



\bibitem{network_coding} R. Ahlswede, N. Cai, S.-Y. R. Li, and R. W. Yeung,
``Network information flow," {\em IEEE Transactions on Information
Theory}, vol. 46, no. 4, pp. 1204-1216, April 2000.

\bibitem{network_coding_coop}Y. Chen, S. Kishore, and J. Li,
``Wireless diversity through network coding," in {\em Proc. IEEE
Wireless Communications and Networking Conference}, Las Vegas, NV,
April 2006.

\bibitem{CISS} Y. Wu, P. A. Chou and S. -Y. Kung, ``Information
exchange in wireless networks with network coding and
physical-layer broadcast," in {\em Proc. 39th Annal Conference on
Information Sciences and Systems}, The Johns Hopkins
University, Baltimore, MD, March 2005.

\bibitem{Laneman} J. N. Laneman, ``Network coding gain of cooperative
diversity," in {\em Proc. IEEE 2004 Military Communications
Conference}, Monterey, CA, October 2004.




\bibitem{Vojcic} Y. Cao and B. Vojcic, ``MMSE multiuser detection
for cooperative diversity CDMA systems," in {\em Proc. IEEE Wireless
Communications and Networking Conference}, Atlanta, GA, March
2004.



\bibitem{Asym1} D. N. C. Tse and S. V. Hanly, ``Linear multiuser receivers: Effective interference,
effective bandwidth and user capacity," {\em IEEE Transactions on
Information Theory}, vol. 45, no. 2, p.p. 641-657, March 1999.

\bibitem{Asym2} T. Tanaka, ``A statistical-mechanics approach to large-system
analysis of CDMA multiuser detectors," {\em IEEE Transactions on
Information Theory}, vol. 48, no. 11, p.p. 2888-2910, November
2002.

\bibitem{Asym3} D. Guo and S. Verd\'{u}, ``Randomly spread CDMA: Asymptotics
via statistical physics," {\em IEEE Transactions on Information
Theory}, vol. 51, no. 6, p.p. 1983-2010, June 2005.


\bibitem{Goldsmith} S.\ Chung and A.\ Goldsmith
\newblock ``Degrees of freedom in adaptive modulation: A unified view",
\newblock {\em IEEE Transactions on Communications}, vol.\ 49, no. 9, pp.\ 1561-1571, September
2001.

\end{thebibliography}





\end{document}